\RequirePackage{fix-cm}
\documentclass[twocolumn,epjc3]{svjour3}  
\usepackage{amsmath}
\usepackage{mathrsfs}
\usepackage{amssymb}
\usepackage{mathtools}
\usepackage{amsmath}
\usepackage{csquotes}
\usepackage{booktabs}
\usepackage{lipsum} %
\smartqed  
\PassOptionsToPackage{colorlinks=true,linkcolor=blue,citecolor=red,urlcolor=magenta}{hyperref}
\usepackage{orcidlink}
\RequirePackage{graphicx}
\usepackage{microtype}
\usepackage{graphicx} %
\usepackage{subcaption} %
\usepackage{stfloats}
\usepackage[numbers,sort&compress]{natbib}

\journalname{Eur. Phys. J. C}
\begin{document}

\title{Covariant Phase Space and Carroll–Weyl $\chi$ Symmetry of Carroll Non-BPS D$_p$-branes}

\author{Limin Zeng\orcidlink{0009-0003-5703-2687}\thanksref{e1,addr1,addr2,addr3}
}

\thankstext{e1}{e-mail: zenglimin25@mails.ucas.ac.cn}


\institute{School of Fundamental Physics and Mathematical Sciences,\\Hangzhou Institute for Advanced
Study, UCAS, Hangzhou 310024, China \label{addr1}
           \and 
           Institute of Theoretical Physics, Chinese Academy of Sciences, Beijing 100190, China \label{addr2}
           \and
           University of Chinese Academy of Sciences, Beijing 100049, China
           \label{addr3}
}

\date{Received: date / Accepted: date}

\maketitle

\begin{abstract}
We analyze the covariant phase space of Kluso\v{n}'s canonical Carroll
non-BPS D\(_p\)-brane actions. The Carroll limit is formulated as a contraction
of the canonical phase space: the canonical one-form is invariant under the
scaling of conjugate pairs, while the leading Hamiltonian constraint differs
between the electric-like and magnetic-like sectors. This difference controls
the weak closure of the electric-like constraint algebra and the stronger
closure of the magnetic-like Hamiltonian brackets. We separate the generic
non-BPS sector, which carries \(D-1\) local phase-space degrees of freedom,
from the tachyon-vacuum sector, which carries \(D-2\) only after imposing
second-class background conditions. We also identify the rank condition for the
generic electric-like sector and its strengthening in the vacuum sector.
Finally, we show that the Carroll--Weyl \(\chi\) transformation, the matter-sector analogue of the null-string Carroll--Weyl symmetry, preserves the symplectic form and admits a charge, but is obstructed by the constraints and cannot be promoted to a first-class gauge generator except in highly restricted corners, a global vacuum sector or the singular potential \(V\propto T^{-1}\) at \(p=1\). The obstruction is algebraic in \(\chi\), so no restriction of its spacetime profile can remove it. This contrasts with null string theory, where
a restricted \(\chi\)-symmetry can be completed to an additional first-class
constraint.

\end{abstract}

\section{Introduction}
\label{sec:1}

Recent progress on flat holography and the asymptotic symmetries of null infinity has triggered a systematic study of Carrollian and tensionless limits of extended objects in string theory \cite{deBoer:2023fnj,Bagchi:2025vri,Ciambelli:2025unn,Argandona:2025jhg,Ruzziconi:2026bix,Bagchi:2026wcu,Blair:2025nno,Ballesteros:2026bqe,Bandos:2026pdg}. Carrollian symmetries, the ultrarelativistic contraction of the Poincaré group, provide the kinematical framework for null hypersurfaces, and their conformal completion was recognized early on as the symmetry of asymptotically flat spacetimes \cite{Duval:2014uva}. In string theory, the tensionless (null) string realizes these symmetries on the worldsheet \cite{Isberg:1993av,Bagchi:2013bga}, and null strings have been proposed as the worldsheet description of strings near black-hole horizons \cite{Bagchi:2023cfp}. A long-standing puzzle concerns the physical counting of the null string. The number of degrees of freedom depends on how the Carroll–Weyl $(\chi)$ symmetry is treated, a discrepancy sharpened in \cite{Sheikh-Jabbari:2026cnj} and systematized in the gauged formulations of the null string \cite{Sheikh-Jabbari:2026vqh,Sheikh-Jabbari:2026tpf}. The quantum theory of these strings has recently been addressed in several complementary ways. The BRST quantization of the Carrollian bosonic string, with both worldsheet and target spacetime Carrollian, was carried out in \cite{Figueroa-OFarrill:2025njv}, where the residual gauge symmetry was shown to form the extended $\text{BMS}_3$ algebra and the spectrum to be finite-dimensional. A systematic comparison of the conformal, ILST, $\chi$ gauged and hybrid null-string formulations shows that the quantum anomalies and critical dimensions depend strongly on the choice of vacuum and normal-ordering prescription \cite{Chen:2026cau}. The path-integral quantization of null strings with Carroll–Weyl ghosts is discussed in \cite{Duary:2026rlo} and the BRST quantization is discussed in \cite{Duary:2026lmk}. Beyond the string, null $p$-branes without a worldvolume gauge sector have been quantized \cite{Dutta:2024gkc,Chen:2026klv}. Given the central role of the Carroll--Weyl symmetry in this debate, it is natural to ask whether Carroll D-branes with worldvolume gauge and tachyon sectors admit an analogous symmetry. The present paper answers this question in the negative for Kluso\v{n}'s canonical Carroll non-BPS D\(_p\)-branes.

Kluso\v{n}'s action predates the gauged Carroll--Weyl formulations, and the null string shows that this order is not essential. There the Carroll--Weyl symmetry is a restricted symmetry of the ILST action \cite{Isberg:1993av}, and the gauging came later \cite{Sheikh-Jabbari:2026vqh,Sheikh-Jabbari:2026tpf}. Testing an existing Carroll action for the analogous symmetry is therefore the procedure of the reference literature. What makes the negative answer informative is that it localizes the obstruction. The obstruction is algebraic in \(\chi\), a connection cannot remove it (Section~\ref{sec:null-string-comparison}), and a completion must introduce a field that carries the compensating weight.

Kluso\v{n}~\cite{Kluson:2017fam} constructed Carroll limits of non-BPS D\(_p\)-branes by starting
from the canonical Hamiltonian form of the unstable D\(_p\)-brane action and
taking different Carroll scalings. The purpose of this
paper is to give a covariant phase-space analysis of these canonical Carroll
non-BPS D\(_p\)-brane actions. We formulate spatial diffeomorphisms, the Gauss
constraint, and the Carroll--Weyl \(\chi\) transformation on the same
presymplectic footing. The Carroll limit is shown to be well defined at the
symplectic level: the canonical one-form is invariant under the scaling of
conjugate pairs, while the electric-like and magnetic-like sectors differ in
the leading part of the Hamiltonian constraint. In particular, we test whether
\(\chi\) is a genuine gauge redundancy or merely a formal grading of the
matter sector, and show that it cannot be promoted to a first-class gauge generator except in highly restricted corners, a global vacuum sector or the singular potential \(V\propto T^{-1}\) at \(p=1\). The same
covariant phase-space framework also makes transparent the weak closure of the
electric-like constraint algebra and the stronger closure properties of the
magnetic-like sector.

The paper is organized as follows. Section~\ref{sec:actions} sets up
the canonical Carroll non-BPS D\(_p\)-brane actions and fixes the
conventions. It also identifies the electric-like and magnetic-like leading
sectors by relating them to the two gauge-field scaling choices in
Kluso\v{n}'s construction, and formulates the Carroll limit as a contraction
of the canonical phase space. Section~\ref{sec:phase-space} introduces the
Carroll presymplectic structure and computes the action of spatial
diffeomorphisms on the canonical variables. We show that the electric momentum
\(\pi^i\) transforms as a vector density only modulo the Gauss constraint.
Section~\ref{sec:constraint-algebra} derives the constraint algebra. The
electric-like algebra closes weakly, whereas the magnetic-like
Hamiltonian--diffeomorphism and Hamiltonian--Hamiltonian brackets close
strongly, although the diffeomorphism--diffeomorp\-hism bracket remains weak.
Section~\ref{sec:degrees} counts the degrees of freedom and discusses the rank
condition. The generic sector carries \(D-1\) local phase-space degrees of
freedom, while the tachyon-vacuum sector carries \(D-2\) only after imposing
second-class background conditions. Section~\ref{sec:chi} analyzes the
Carroll--Weyl \(\chi\) transformation, states the resulting no-go statement,
and compares the role of \(\chi\) in the Carroll D\(_p\)-brane with its role in
null string theory. Section~\ref{sec:conclusion} summarizes our conclusions. Appendix~\ref{app:A} collects the derivation of the diffeomorphism
transformation of \(\pi^i\).

\section{Canonical Carroll non-BPS D$_p$-brane actions and leading sectors}
\label{sec:actions}

\subsection{Parent theory}
\label{sec:variables-conventions}
We start from the non-BPS D$_p$ brane action: 
\begin{equation}
\begin{split}
S &= -\tilde{\tau}_p \int d^{p+1}\xi \, V(\tilde{T}) \sqrt{-\det \mathcal{A}},\\ \mathcal{A}_{\alpha\beta} &= g_{MN} \partial_\alpha \tilde{x}^M \partial_\beta \tilde{x}^N + \tilde{F}_{\alpha\beta} + \partial_\alpha \tilde{T} \partial_\beta \tilde{T}.
\end{split}
\tag{2.1} \label{eq:2.1}
\end{equation}
The canonical version of \eqref{eq:2.1} is
\begin{equation}
\begin{aligned}
  S &= \int d^{p+1}\xi \Big( \tilde{p}_M \partial_0 \tilde{x}^M + \tilde{p}_T \partial_0 \tilde{T} + \tilde{\pi}^i \partial_0 \tilde{A}_i \\
    &\qquad - \tilde{N} \widetilde{\mathcal H}_\tau - \tilde{N}^i \widetilde{\mathcal H}_i - \tilde{\pi}^i \partial_i \tilde{A}_0 \Big), \\[8pt]
  \widetilde{\mathcal H}_\tau &= \tilde{p}^M g_{MN} \tilde{p}^N + \tilde{p}_T^2 + \tilde{\pi}^i \mathcal A^S_{ij} \tilde{\pi}^j + \tilde{\tau}_p^2 V^2 \det \mathcal A_{ij} \approx 0, \\[8pt]
\widetilde{\mathcal H}_i &= \tilde{p}_M \partial_i \tilde{x}^M + \tilde{p}_T \partial_i \tilde{T} + \tilde{F}_{ij} \tilde{\pi}^j.
\end{aligned}
\tag{2.2} \label{eq:2.2}
\end{equation}
Indices are raised and lowered with \(g^{MN}\) and \(g_{MN}\).
The worldvolume coordinates of the non-BPS D$_p$ brane are
\(
\xi^\alpha=(\tau,\sigma^i),
\quad
i=1,\ldots,p.
\)
The target-space coordinates are split into longitudinal and transverse
sets,
\begin{equation}
\tilde x^M=(\tilde x^\mu,\tilde x^I),
\quad
\mu=0,\ldots,p,
\quad
I=p+1,\ldots,D-1.
\tag{2.3} \label{eq:2.3}
\end{equation}
The non-BPS D$p$-brane contains a tachyon \(\tilde T\) and a worldvolume gauge field \(\tilde A_\alpha\) \cite{Sen:1999md} and the tachyon potential and its vacuum structure are reviewed in Ref.~\cite{Sen:2004nf}.
The canonical phase space $\widetilde\Gamma$ is parametrized by the pairs
\begin{equation}
(\tilde x^\mu,\tilde p_\mu),
\quad
(\tilde x^I,\tilde p_I),
\quad
(\tilde T,\tilde p_T),
\quad
(\tilde A_i,\tilde \pi^i).
\tag{2.4} \label{eq:2.4}
\end{equation}
The spatial field strength and the transverse induced metric are
\begin{equation}
\tilde F_{ij}=\partial_i \tilde A_j-\partial_j \tilde A_i,
\qquad
\tilde a_{ij}
=
\partial_i \tilde x^I\partial_j \tilde x^I
+
\partial_i \tilde T\,\partial_j \tilde T.
\tag{2.5} \label{eq:2.5}
\end{equation}
For the longitudinal target-space metric we adopt the mostly-plus
convention
\begin{equation}
\eta^{\mu\nu}=\operatorname{diag}(-1,+1,\ldots,+1),
\qquad
\mu,\nu=0,\ldots,p,
\tag{2.6} \label{eq:2.6}
\end{equation}
so that
\begin{equation}
\tilde p_\mu\eta^{\mu\nu}\tilde p_\nu
=
-\tilde p_0^2+\sum_{j=1}^{p}\tilde p_j^2.
\tag{2.7} \label{eq:2.7}
\end{equation}
All Hamiltonian constraints below are written in this convention.

The parent theory is most conveniently formulated in covariant phase-space
language. Let \(\Sigma\) be a spatial slice with coordinates \(\sigma^i\).
The canonical one-form on the parent phase space is
\begin{equation}
\widetilde\Theta
=
\int_\Sigma
\left(
\tilde p_M\,\delta\tilde x^M
+
\tilde p_T\,\delta\tilde T
+
\tilde\pi^i\,\delta\tilde A_i
\right).
\tag{2.8} \label{eq:2.8}
\end{equation}
The corresponding presymplectic form is
\begin{equation}
\widetilde\Omega
=
\delta\widetilde\Theta
=
\int_\Sigma
\left(
\delta\tilde p_M\wedge\delta\tilde x^M
+
\delta\tilde p_T\wedge\delta\tilde T
+
\delta\tilde\pi^i\wedge\delta\tilde A_i
\right).
\tag{2.9}\label{eq:2.9}
\end{equation}
Equivalently, the canonical Poisson brackets are
\begin{equation}
\begin{split}
\{\tilde x^M(\sigma),\tilde p_N(\sigma')\}
&=
\delta^M{}_N\,\delta^p(\sigma-\sigma'),\\
\{\tilde T(\sigma),\tilde p_T(\sigma')\}
&=
\delta^p(\sigma-\sigma'),
\\
\{\tilde A_i(\sigma),\tilde\pi^j(\sigma')\}
&=
\delta_i{}^j\,\delta^p(\sigma-\sigma').
\end{split}
\tag{2.10}\label{eq:2.10}
\end{equation}

The total Hamiltonian is a linear combination of constraints,
\begin{equation}
\widetilde H_{\rm tot}
=
\int_\Sigma
\left(
\tilde N\,\widetilde{\mathcal H}_\tau
+
\tilde N^i\,\widetilde{\mathcal H}_i
+
\tilde A_0\,\widetilde{\mathcal G}
\right),
\tag{2.11} \label{eq:2.11}
\end{equation}
where
\begin{equation}
\widetilde{\mathcal G}
=
\partial_i\tilde\pi^i
\approx0.
\tag{2.12} \label{eq:2.12}
\end{equation}
Thus the dynamics is entirely constrained. The physical phase space of the
parent theory is obtained by restricting to the constraint surface
\begin{equation}
\widetilde{\mathcal C}
=
\left\{
\widetilde{\mathcal H}_\tau\approx0,\quad
\widetilde{\mathcal H}_i\approx0,\quad
\widetilde{\mathcal G}\approx0
\right\},
\tag{2.13}\label{eq:2.13}
\end{equation}
and quotienting by gauge transformations generated by the first-class constraints.

It is useful to introduce the smeared constraints
\begin{equation}
\widetilde G[\alpha]
=
\int_\Sigma
\alpha\,\widetilde{\mathcal G},
\quad
\widetilde H_i[\xi]
=
\int_\Sigma
\xi^i\,\widetilde{\mathcal H}_i,
\quad
\widetilde H_\tau[\lambda]
=
\int_\Sigma
\lambda\,\widetilde{\mathcal H}_\tau.
\tag{2.14} \label{eq:2.14}
\end{equation}
These generate the gauge transformations of the parent D\(p\)-brane.
What matters for the Carroll contraction is that the parent phase space is
equipped with \(\widetilde\Omega\) and the constraint surface
\(\widetilde{\mathcal C}\). In the next subsection we implement the Carroll scaling of the canonical
pairs and show how \(\widetilde\Omega\) and \(\widetilde{\mathcal C}\) reduce
to their Carroll limits.

\subsection{Carroll limit}
\label{sec:two-sectors}
One rescaling choice of the D$_p$ brane theory is 
\begin{equation}
\begin{aligned}
  \tilde{x}^\mu &= \frac{X^\mu}{\omega}, & \quad \tilde{p}_\mu &= \omega P_\mu, \\
  \tilde{x}^I &= X^I, & \quad \tilde{p}_I &= P_I, \\
  \tilde{A}_i &= \frac{A_i}{\omega}, & \quad \tilde{\pi}^i &= \omega \pi^i, \\
  \tilde{A}_0 &= \frac{1}{\omega} A_0, & \quad \tilde{N} &= \frac{1}{\omega^2} N, \\
  \tilde{N}^i &= N^i, & \quad \tilde{\tau}_p &= \omega \tau_p, \\
  \tilde{T} &= T, & \quad \tilde{p}_T &= p_T.
\end{aligned}
\tag{2.15} \label{eq:2.15}
\end{equation}
Substituting this scaling into \eqref{eq:2.8}, we obtain
\begin{equation}
\Theta_E^{\rm Car}
=
\int_\Sigma
\left(
P_\mu\delta X^\mu
+
P_I\delta X^I
+
p_T\delta T
+
\pi^i\delta A_i
\right)=\tilde\Theta.
\tag{2.16} \label{eq:2.16}
\end{equation}
We take the limit $\omega\to \infty$, then $
\widetilde{\mathcal H}_\tau=\omega^2 h_E+\mathcal{O}(1)$, where 
\begin{equation}
\begin{split}
h_E
&=P_\mu\eta^{\mu\nu}P_\nu
+\pi^i a_{ij}\pi^j
+\tau_p^2V(T)^2\det a_{ij}\approx0,
\\a_{ij} &= \partial_i X^I \partial_j X^I + \partial_i T \partial_j T.
\end{split}
\tag{2.17} \label{eq:2.17}
\end{equation}
Since multiplying a constraint by a nonzero function does not change the
constraint surface, we define the rescaled constraint
\(
\mathcal H_\omega
:=
\omega^{-2}\widetilde{\mathcal H}_\tau.
\)
Then
\(
\mathcal H_\omega\to h_E.
\)
Equivalently, the factor \(\omega^{-2}\) can be absorbed into the lapse
\(\tilde N=N/\omega^2\). The rescaling is therefore only a normalization choice needed to obtain a finite Carroll Hamiltonian constraint and it is not an independent scaling of the constraint.
The resulting Carroll action is
\begin{equation}
\begin{split}
  S_E = \int d\tau\,d^p\sigma \Bigl[ 
    & P_\mu\partial_\tau X^\mu + P_I\partial_\tau X^I + p_T\partial_\tau T + \pi^i\partial_\tau A_i \\
    & - \pi^i\partial_i A_0 - N h_E - N^i\mathcal{H}_i \Bigr].
\end{split}
\tag{2.18} \label{eq:2.18}
\end{equation}
We denote this theory as the electric-like theory. The specific reasons are discussed below.

We can also choose the magnetic-like scaling as follows,
\begin{equation}
\begin{aligned}
  \tilde{x}^\mu &= \frac{X^\mu}{\omega}, & \quad \tilde{p}_\mu &= \omega P_\mu, \\
  \tilde{x}^I &= X^I, & \quad \tilde{p}_I &= P_I, \\
  \tilde{T} &= T, & \quad \tilde{p}_T &= p_T, \\
  \tilde{A}_i &= A_i, & \quad \tilde{\pi}^i &= \pi^i, \\
  \tilde{A}_0 &= A_0, & \quad \tilde{N} &= \frac{N}{\omega^2}, & \quad \tilde{N}^i &= N^i, \\
  \tilde{\tau}_p &= \omega \tau_p.
\end{aligned}
\tag{2.19} \label{eq:2.19}
\end{equation}
Similarly, the time reparametrization constraint becomes
\begin{equation}
\begin{split}
h_M
&=
P_\mu\eta^{\mu\nu}P_\nu
+
\tau_p^2V(T)^2\det(b_{ij})\approx0,\\b_{ij}&=a_{ij}+F_{ij}.
\end{split}
\tag{2.20} \label{eq:2.20}
\end{equation}
Substituting scalings \eqref{eq:2.19} into \eqref{eq:2.8}, we can also obtain
\begin{equation}
\Theta_M^{\rm Car}
=
\int_\Sigma
\left(
P_\mu\delta X^\mu
+
P_I\delta X^I
+
p_T\delta T
+
\pi^i\delta A_i
\right)=\tilde\Theta.
\tag{2.21} \label{eq:2.21}
\end{equation}
In both the electric-like and magnetic-like Carroll scalings, the canonical one-form is invariant under the scaling of conjugate pairs, so the Carroll limit is well defined at the symplectic level. The difference between the two sectors lies not in the presymplectic structure but in the leading scaling of the Hamiltonian constraint.
And the canonical action of magnetic-like theory is 
\begin{equation}
\begin{split}
  S_M = \int d^{p+1}\xi \Bigl[ 
    & P_\mu\partial_\tau X^\mu + P_I\partial_\tau X^I + p_T\partial_\tau T + \pi^i\partial_\tau A_i \\
    & - \pi^i\partial_i A_0 - N h_M - N^i\mathcal{H}_i \Bigr].
\end{split}
\tag{2.22} \label{eq:2.22}
\end{equation}
We denote this theory as the magnetic-like theory.
However, \eqref{eq:2.18} and \eqref{eq:2.22} share the same Gauss and spatial diffeomorphism constraints.
The Gauss constraint is
\begin{equation}
\mathcal G=\partial_i\pi^i\approx0.
\tag{2.23} \label{eq:2.23}
\end{equation}
The spatial diffeomorphism constraint is
\begin{equation}
\mathcal H_i
=
P_\mu\partial_i X^\mu
+
P_I\partial_i X^I
+
p_T\partial_i T
+
F_{ij}\pi^j
\approx0.
\tag{2.24} \label{eq:2.24}
\end{equation}
The temporal component \(A_0\) appears in the action only through the
combination
\(
-\pi^i\partial_i A_0
=
A_0\,\partial_i\pi^i
\)
up to boundary terms, so it is the multiplier of the Gauss constraint. We use the terms electric-like and magnetic-like to denote the two leading Carroll sectors of Kluso\v{n}'s construction, distinguished by whether the leading Hamiltonian density contains the electric flux term $\pi^i a_{ij}\pi^j$ or the magnetic DBI determinant $\det(a_{ij}+F_{ij})$, respectively.

\subsection{Density weights}
\label{sec:density-weights}
Before proceeding to the covariant phase-space analysis, it is useful to
record the density weights of the canonical variables, the multipliers, and
the constraint densities. These weights determine how the Carroll action
transforms under spatial diffeomorphisms and will be used to discuss the
weak/strong distinction later.
The elementary phase-space variables carry the weights listed in
Table~\ref{tab:weights}. The weights of \(h_E\) and \(h_M\) are formal
tensor-density weights and their transformation laws under the spatial
diffeomorphism constraint are derived in Section~\ref{sec:constraint-algebra}.
Since \(h_E\) and \(h_M\) carry density weight \(+2\), while \(N\) has
weight \(-1\), the products \(N h_E\) and \(N h_M\) are weight-one densities.
The kinetic terms
\[
P_\mu\partial_\tau X^\mu,
\qquad
P_I\partial_\tau X^I,
\qquad
p_T\partial_\tau T,
\qquad
\pi^i\partial_\tau A_i
\]
are also weight-one densities. With these conventions fixed, we can now formulate the Carroll phase space
and analyze the action of spatial diffeomorphisms on it.
\begin{table}[htbp]
\centering
\begin{tabular}{@{}ll@{}}
\toprule
variable & density weight \\
\midrule
\(X^\mu,\ X^I,\ T\) & \(0\) \\
\(A_i\) & \(0\) (covector) \\
\(P_\mu,\ P_I,\ p_T,\ \pi^i\) & \(+1\) \\
\(N\) & \(-1\) \\
\(N^i\) & \(0\) (vector) \\
\(A_0\) & \(0\) (scalar) \\
\midrule
\(h_M\) & \(+2\) \\
\(h_E\) & \(+2\) \\
\(\mathcal H_i\) & \(+1\) \\
\(\mathcal G\) & \(+1\) \\
\bottomrule
\end{tabular}
\caption{Formal density weights of the canonical variables, the multipliers,
the Hamiltonian densities, and the constraint densities on the spatial slice
\(\Sigma\).}
\label{tab:weights}
\end{table}

\section{Covariant phase space and gauge transformation}
\label{sec:phase-space}

\subsection{Presymplectic structure}
\label{sec:presymplectic}

The Carroll presymplectic potential is
\begin{equation}
\Theta
=
\int_\Sigma
\Bigl(
P_\mu\,\delta X^\mu
+
P_I\,\delta X^I
+
p_T\,\delta T
+
\pi^i\,\delta A_i
\Bigr),
\tag{3.1} \label{eq:3.1}
\end{equation}
and the presymplectic form is
\begin{equation}
\Omega
=
\int_\Sigma
\Bigl(
\delta P_\mu\wedge\delta X^\mu
+
\delta P_I\wedge\delta X^I
+
\delta p_T\wedge\delta T
+
\delta\pi^i\wedge\delta A_i
\Bigr).
\tag{3.2} \label{eq:3.2}
\end{equation}

The smeared diffeomorphism constraint is
\begin{equation}
H_i[\xi]
=
\int_\Sigma
\xi^i
\Bigl(
P_\mu\partial_i X^\mu
+
P_I\partial_i X^I
+
p_T\partial_i T
+
F_{ij}\pi^j
\Bigr),
\tag{3.3} \label{eq:3.3}
\end{equation}
the smeared Gauss constraint is
\begin{equation}
G[\alpha]
=
\int_\Sigma
\alpha\,\partial_i\pi^i,
\tag{3.4} \label{eq:3.4}
\end{equation}
and the smeared Hamiltonian constraints are
\begin{equation}
H_E[\lambda]
=
\int_\Sigma \lambda h_E,
\qquad
H_M[\lambda]
=
\int_\Sigma \lambda h_M.
\tag{3.5} \label{eq:3.5}
\end{equation}
With the convention
\begin{equation}
\delta_G F=\{F,G\},
\tag{3.6} \label{eq:3.6}
\end{equation}
these generate the gauge transformations of the Carroll theory.
\subsection{Gauss transformation}
The Gauss constraint generates the \(U(1)\) gauge transformation
\begin{equation}
\delta_\alpha A_i
=
\{A_i,G[\alpha]\}
=
-\partial_i\alpha,
\qquad
\delta_\alpha\pi^i=0.
\tag{3.7} \label{eq:3.7}
\end{equation}
\subsection{Spatial diffeomorphism transformation}
The smeared diffeomorphism constraint $H_i[\xi]$ generates spatial diffeomorphisms on
the matter variables:
\begin{equation}
\begin{aligned}
  \delta_\xi X^\mu &= \xi^i\partial_iX^\mu, & \quad \delta_\xi P_\mu &= \mathcal L_\xi P_\mu, \\[6pt]
  \delta_\xi X^I &= \xi^i\partial_iX^I, & \quad \delta_\xi P_I &= \mathcal L_\xi P_I, \\[6pt]
  \delta_\xi T &= \xi^i\partial_iT, & \quad \delta_\xi p_T &= \mathcal L_\xi p_T.
\end{aligned}
\tag{3.8} \label{eq:3.8}
\end{equation}
On the gauge sector it acts as a diffeomorphism plus a field-dependent gauge
transformation:
\begin{equation}
\delta_\xi A_i
=
\{A_i,H_j[\xi]\}
=
\mathcal L_\xi A_i
-
\partial_i(\xi^jA_j).
\tag{3.9} \label{eq:3.9}
\end{equation}
For the gauge-field momentum one obtains
\begin{equation}
\delta_\xi\pi^i
=
\{\pi^i,H_j[\xi]\}
=
\mathcal L_\xi\pi^i
-
\xi^i\mathcal G.
\tag{3.10} \label{eq:3.10}
\end{equation}
Thus \(\pi^i\) transforms as a vector density only modulo the Gauss
constraint.
\subsection{Hamiltonian transformation}
The Hamiltonian constraint generates Carroll time repa\-rametrizations. For the
electric-like Hamiltonian generator $H_E[\lambda]$ with smeared parameter $\lambda$,
the nontrivial transformations are
\begin{equation}
\begin{aligned}
  \delta_\lambda X^\mu &= \{X^\mu, H_E[\lambda]\} = 2\lambda\eta^{\mu\nu}P_\nu, \\[6pt]
  \delta_\lambda A_i &= \{A_i, H_E[\lambda]\} = 2\lambda a_{ij}\pi^j, \\[6pt]
  \delta_\lambda P_I &= -\frac{\delta H_E[\lambda]}{\delta X^I}, \qquad 
  \delta_\lambda p_T = -\frac{\delta H_E[\lambda]}{\delta T}.
\end{aligned}
\tag{3.11}\label{eq:3.11}
\end{equation}
The remaining variables are inert under the electric Hamiltonian constraint:
\begin{equation}
\delta_\lambda X^I=0,
\qquad
\delta_\lambda T=0,
\qquad
\delta_\lambda P_\mu=0,
\qquad
\delta_\lambda\pi^i=0.
\tag{3.12}\label{eq:3.12}
\end{equation}
For the magnetic-like Hamiltonian generator $H_M[\lambda]$ one has
\begin{equation}
\begin{aligned}
  \delta_\lambda X^\mu &= 2\lambda\eta^{\mu\nu}P_\nu, \\[6pt]
  \delta_\lambda A_i &= 0, \\[6pt]
  \delta_\lambda\pi^i
&=
\partial_j
\left[
\lambda\tau_p^2V(T)^2\det b
\left(
(b^{-1})^{ji}-(b^{-1})^{ij}
\right)
\right], \\[6pt]
  \delta_\lambda P_I &= -\frac{\delta H_M[\lambda]}{\delta X^I}, \qquad 
  \delta_\lambda p_T = -\frac{\delta H_M[\lambda]}{\delta T}.
\end{aligned}
\tag{3.13}\label{eq:3.13}
\end{equation}
The remaining variables are inert:
\begin{equation}
\delta_\lambda X^I=0,
\qquad
\delta_\lambda T=0,
\qquad
\delta_\lambda P_\mu=0.
\tag{3.14}\label{eq:3.14}
\end{equation}
The important point is that the electric-like Hamiltonian flow acts on
\(A_i\) through the electric momentum \(\pi^i\), whereas the magnetic-like
Hamiltonian flow acts on \(\pi^i\) through the field strength \(F_{ij}\).
In both sectors, however, the leading Hamiltonian constraint does not contain
\(P_I^2\) or \(p_T^2\). Therefore the transverse scalars and the tachyon do
not evolve under the Hamiltonian constraint at leading order.

\section{Constraint algebra}
\label{sec:constraint-algebra}

\subsection{Electric-like sector: weak closure}
\label{sec:electric-closure}

The electric Hamiltonian density is $h_E$ \eqref{eq:2.17}.
The first and third terms in \eqref{eq:2.17} transform as weight-two
densities strongly. The flux term, by contrast, transforms weakly:
\begin{equation}
\{h_E,H_i[\xi]\}
=
\mathcal L_\xi h_E
-
2\xi^i\mathcal G\,a_{ij}\pi^j.
\tag{4.1}\label{eq:4.1}
\end{equation}

For the smeared Hamiltonian constraint $H_E[\lambda]$, one obtains
\begin{equation}
\{H_i[\xi],H_E[\lambda]\}
=
H_E[\mathcal L_\xi\lambda]
+
2\int_\Sigma
\lambda\,\xi^i\mathcal G\,a_{ij}\pi^j.
\tag{4.2}\label{eq:4.2}
\end{equation}
On the constraint surface the second term vanishes, so that
\begin{equation}
\{H_i[\xi],H_E[\lambda]\}
\approx
H_E[\mathcal L_\xi\lambda].
\tag{4.3}\label{eq:4.3}
\end{equation}

The electric Hamiltonian constraint commutes strongly with itself,
\begin{equation}
\{H_E[\lambda],H_E[\mu]\}=0,
\tag{4.4}\label{eq:4.4}
\end{equation}
because in every canonical pair at least one of the two functional derivatives of \(h_E\) vanishes. The density \(h_E\) is independent of \(X^\mu\), of \(P_I\), of \(p_T\), and of \(A_i\). It does depend on \(X^I\) and \(T\), through \(a_{ij}\) and \(V(T)\), but these variables are paired in the Poisson bracket with the absent momenta \(P_I\) and \(p_T\).
It also commutes strongly with the Gauss constraint,
\begin{equation}
\{G[\alpha],H_E[\lambda]\}=0,
\tag{4.5}\label{eq:4.5}
\end{equation}
because \(h_E\) does not depend on \(A_i\).

\subsection{Magnetic-like sector: stronger closure}
\label{sec:magnetic-closure}

The magnetic Hamiltonian density is $h_M$ \eqref{eq:2.20}.
It does not depend on \(\pi^i\). Since \(F_{ij}\) is a tensor,
\begin{equation}
\{h_M,H_i[\xi]\}
=
\mathcal L_\xi h_M,
\tag{4.6}\label{eq:4.6}
\end{equation}
and therefore
\begin{equation}
\{H_i[\xi],H_M[\lambda]\}
=
H_M[\mathcal L_\xi\lambda],
\tag{4.7}\label{eq:4.7}
\end{equation}
strongly. The remaining brackets are also strong:
\begin{equation}
\{H_M[\lambda],H_M[\mu]\}=0,
\qquad
\{G[\alpha],H_M[\lambda]\}=0.
\tag{4.8}\label{eq:4.8}
\end{equation}

\subsection{Diffeomorphism--diffeomorphism bracket}
\label{sec:diff-diff}

Split the diffeomorphism density into its matter and gauge parts,
\begin{equation}
\mathcal H_i=\mathcal H_i^{\mathrm m}+\mathcal H_i^{\mathrm g},
\tag{4.9}\label{eq:4.9}
\end{equation}
where
\begin{equation}
\mathcal H_i^{\mathrm m}
=
P_\mu\partial_i X^\mu
+
P_I\partial_i X^I
+
p_T\partial_i T,
\qquad
\mathcal H_i^{\mathrm g}=F_{ij}\pi^j.
\tag{4.10}\label{eq:4.10}
\end{equation}
The matter part transforms strongly,
\begin{equation}
\{\mathcal H_i^{\mathrm m},H_k[\xi]\}
=
\mathcal L_\xi\mathcal H_i^{\mathrm m},
\tag{4.11}\label{eq:4.11}
\end{equation}
whereas the gauge part picks up a Gauss term,
\begin{equation}
\{\mathcal H_i^{\mathrm g},H_k[\xi]\}
=
\mathcal L_\xi\mathcal H_i^{\mathrm g}
-
\xi^jF_{ij}\mathcal G.
\tag{4.12}\label{eq:4.12}
\end{equation}
Combining the two contributions,
\begin{equation}
\{\mathcal H_i,H_k[\xi]\}
=
\mathcal L_\xi\mathcal H_i
-
\xi^jF_{ij}\mathcal G.
\tag{4.13}\label{eq:4.13}
\end{equation}
A direct computation then gives
\begin{equation}
\{H_i[\xi],H_i[\eta]\}
=
H_i[\mathcal L_\xi\eta]
-
\frac12\int_\Sigma
(\xi^i\eta^j-\eta^i\xi^j)F_{ij}\,\mathcal G.
\tag{4.14}\label{eq:4.14}
\end{equation}
On the constraint surface,
\begin{equation}
\{H_i[\xi],H_i[\eta]\}
\approx
H_i[\mathcal L_\xi\eta].
\tag{4.15}\label{eq:4.15}
\end{equation}

\subsection{Diffeomorphism--Gauss bracket}
\label{sec:diff-Gauss}

Using \eqref{eq:3.10}, we find
\begin{equation}
\{\mathcal G,H_k[\xi]\}
=
\partial_i(\mathcal L_\xi\pi^i)
-
\partial_i(\xi^i\mathcal G)
=
0.
\tag{4.16}\label{eq:4.16}
\end{equation}
Thus, for fixed smearing $\alpha$,
\begin{equation}
\{G[\alpha],H_i[\xi]\}=0.
\tag{4.17}\label{eq:4.17}
\end{equation}
Table \ref{tab:algebra} summarizes the results above.
\begin{table}[htbp]
\centering
\begin{tabular}{@{}lll@{}}
\toprule
bracket & electric-like & magnetic-like \\
\midrule
\(\{H_i[\xi],H_i[\eta]\}\) &
\(\approx H_i[\mathcal L_\xi\eta]\) &
\(\approx H_i[\mathcal L_\xi\eta]\) \\[2pt]
\(\{H_i[\xi],G[\alpha]\}\) &
\(=0\) &
\(=0\) \\[2pt]
\(\{G[\alpha],G[\beta]\}\) &
\(=0\) &
\(=0\) \\[2pt]
\(\{H_i[\xi],H[\lambda]\}\) &
\(\approx H_E[\mathcal L_\xi\lambda]\) &
\(=H_M[\mathcal L_\xi\lambda]\) \\[2pt]
\(\{H[\lambda],H[\mu]\}\) &
\(=0\) &
\(=0\) \\[2pt]
\(\{G[\alpha],H[\lambda]\}\) &
\(=0\) &
\(=0\) \\
\bottomrule
\end{tabular}
\caption{Constraint algebra in the electric-like and magnetic-like sectors.
The Hamiltonian constraint is \(H_E\) in the electric-like column and
\(H_M\) in the magnetic-like column.}
\label{tab:algebra}
\end{table}

\section{Degrees of freedom and rank condition}
\label{sec:degrees}

\subsection{Generic Dirac count}
\label{sec:generic-count}

The canonical pairs are
\begin{equation}
(X^\mu,P_\mu),
\quad
(X^I,P_I),
\quad
(T,p_T),
\quad
(A_i,\pi^i),
\tag{5.1}\label{eq:5.1}
\end{equation}
so their number is
\begin{equation}
N_{\mathrm{pairs}}
=
(p+1)+(D-p-1)+1+p
=
D+p+1.
\tag{5.2}\label{eq:5.2}
\end{equation}
The first-class constraints are
\begin{equation}
H[\lambda],
\qquad
H_i[\xi^i],
\qquad
G[\alpha],
\tag{5.3}\label{eq:5.3}
\end{equation}
whose number is
\begin{equation}
F=1+p+1=p+2.
\tag{5.4}\label{eq:5.4}
\end{equation}
The generic non-BPS Carroll D$p$-brane therefore has
\begin{equation}
N_{\mathrm{phys}}^{\mathrm{generic}}
=
\frac{1}{2}\left[2(D+p+1)-2(p+2)\right]
=
D-1
\tag{5.5}\label{eq:5.5}
\end{equation}
physical degrees of freedom. This is the result of the first-class Dirac
constraint analysis alone.
As throughout the paper, the count is made on the non-degenerate locus discussed in Section~\ref{sec:rank-condition}. On degenerate loci, where \(\det a_{ij}=0\) and \(\det b_{ij}=0\), a separate analysis is required. Those cases are not claimed here.

\subsection{Tachyon vacuum as an effective sector}
\label{sec:tachyon-vacuum}

The leading Carroll Hamiltonian densities \(h_E\) and \(h_M\) do not contain
\(p_T^2\). Consequently, the first-class constraints do not dynamically force
\(p_T=0\). Likewise, the leading Hamiltonian constraint does not by itself
fix \(T=T_{\min}\). Although the diffeomorphism constraint contains
\(p_T\partial_iT\), this does not imply \(p_T=0\) unless one restricts to
configurations of fixed \(T\).

The reduction from \(D-1\) to \(D-2\) at the tachyon vacuum is therefore an
additional sector selection. We impose the vacuum-sector conditions
\begin{equation}
T-T_{\min}\approx0,
\qquad
p_T\approx0.
\tag{5.6}
\label{eq:vacuum-sector}
\end{equation}
These two conditions have nonzero Poisson bracket,
\begin{equation}
\{T,p_T\}=1,
\tag{5.7}
\label{eq:vacuum-bracket}
\end{equation}
and hence form a second-class pair.  They are not generated by the original first-class constraint algebra, but instead define an effective tachyon vacuum sector. The sector is dynamically consistent. It is preserved by the Hamiltonian flow of both leading constraints. Indeed \(\dot T=\{T,H_{E/M}[\lambda]\}=\delta H_{E/M}[\lambda]/\delta p_T=0\), since neither \(h_E\) nor \(h_M\) contains \(p_T\). At \(T=T_{\min}\), where \(\partial_iT=0\) and \(V=V'=0\), every term of \(\delta h_{E/M}/\delta T\) carries a factor of \(V\), of \(V'\) or of \(\partial_iT\). Hence \(\dot p_T\approx0\) on the sector. The second-class pair \eqref{eq:vacuum-sector} therefore defines a legitimate invariant sector. In this sector one should use the Dirac bracket associated
with the pair \eqref{eq:vacuum-sector}. The phase-space dimension is lowered
by two, corresponding to one canonical pair. The effective vacuum-sector
count is therefore
\begin{equation}
N_{\mathrm{phys}}^{\mathrm{vac}}
=
(D-1)-1
=
D-2.
\tag{5.8}
\label{eq:vacuum-count}
\end{equation}

Equivalently, the vacuum sector is defined by
\begin{equation}
T=T_{\min},
\quad
p_T=0,
\quad
V(T_{\min})=0,
\quad
V'(T_{\min})=0.
\tag{5.9}
\label{eq:vacuum-conditions}
\end{equation}
Under these conditions the tachyon-dependent terms drop out of the leading
Hamiltonian constraints, and the remaining phase space has the same effective
count as a Carroll D\(p\)-brane without a tachyon.

The count \(D-1\) is the generic Dirac count for Kluso\v{n}'s non-BPS Carroll D\(p\)-brane. The count \(D-2\) is the count of the tachyon-vacuum effective sector obtained after imposing the second-class conditions \eqref{eq:vacuum-sector}. It is not an automatic consequence of the first-class constraint algebra.

The count \(D-2\) is a reduced phase-space dimension, not a statement that
every remaining variable propagates in Carrollian time. Frozen
\(\tau\)-evolution does not remove a variable from the reduced phase space.
In particular, in both sectors the leading Hamiltonian constraint does not
contain \(P_I^2\) or \(p_T^2\). The transverse scalars and the tachyon
therefore do not evolve under the Hamiltonian constraint at leading order.
This is a consequence of the Carroll scaling used here, not a universal
property of all Carroll D\(_p\)-brane models.

\subsection{Rank condition}
\label{sec:rank-condition}

In the electric-like sector, \(a_{ij}\) is the spatial metric induced on the
D\(p\)-brane by the transverse scalars and the tachyon. It is a \(p\times p\)
Gram matrix built from
\[
(D-p-1)+1=D-p
\]
transverse vectors, so that
\begin{equation}
\operatorname{rank}a_{ij}\le D-p.
\tag{5.10}
\end{equation}
The generic nondegenerate electric-like sector requires
\(
\operatorname{rank}a_{ij}=p.
\)
This means that the brane still wraps a genuine \(p\)-dimensional spatial
volume, so that \(\det a_{ij}\neq0\) and the DBI volume term has its usual
geometric meaning. For a generic embedding this requires
\begin{equation}
D\ge2p.
\tag{5.11}
\end{equation}
For \(D=10\), this includes \(p\le5\). For \(D<2p\), the generic
electric-like sector is degenerate, and that degenerate sector is not
analyzed here.

In the tachyon-vacuum effective sector one has
\(
\partial_iT=0,
\)
so that \(a_{ij}\) is built only from \(D-p-1\) transverse scalars. The
nondegeneracy condition then becomes
\begin{equation}
D\ge2p+1.
\tag{5.12}
\end{equation}
To summarize, nondegeneracy of \(a_{ij}\) requires \(D\ge2p\) in the generic
sector with a dynamical tachyon, and \(D\ge2p+1\) in the tachyon-vacuum
effective sector.

The magnetic-like sector is not subject to the same restriction, because
\(b_{ij}=a_{ij}+F_{ij}\) can be nondegenerate even when \(a_{ij}\) is
degenerate. The explicit magnetic-sector counterexample of
Section~\ref{sec:magnetic-counterexample} illustrates this mechanism:
\(a_{ij}\) has rank \(p-1\), while \(b_{ij}=a_{ij}+F_{ij}\) is
nondegenerate for \(f\neq0\). In this case the
worldvolume field strength contributes directly to the spatial geometry. A related role of nondegenerate worldvolume field strengths
appears in tensionless higher-brane constructions \cite{Bandos:2026pdg}.

\section{Carroll--Weyl \(\chi\) structure and no-go statement}
\label{sec:chi}

\subsection{Definition and charge}
\label{sec:chi-definition}

We introduce a field-space transformation 
\(\delta_\chi\) that scales the transverse coordinates, the tachyon, and the gauge field as
\begin{equation}
\delta_\chi X^I=\chi X^I,
\qquad
\delta_\chi T=\chi T,
\qquad
\delta_\chi A_i=\chi A_i,
\tag{6.1}\label{eq:6.1}
\end{equation}
while their conjugate momenta transform with the opposite sign,
\begin{equation}
\delta_\chi P_I=-\chi P_I,
\quad
\delta_\chi p_T=-\chi p_T,
\quad
\delta_\chi\pi^i=-\chi\pi^i.
\tag{6.2}\label{eq:6.2}
\end{equation}
The longitudinal variables are inert,
\begin{equation}
\delta_\chi X^\mu=0,
\qquad
\delta_\chi P_\mu=0.
\tag{6.3}\label{eq:6.3}
\end{equation}

The charge that generates these transformations is
\begin{equation}
Q_\chi
=
\int_\Sigma
\chi
\Bigl(
P_I X^I
+
p_T T
+
\pi^i A_i
\Bigr),
\tag{6.4}\label{eq:6.4}
\end{equation}
with the convention
\(
\delta_{Q_\chi}F=\{F,Q_\chi\}.
\)
For each canonical pair, the coordinate and the momentum scale with opposite
weights. Here \(\chi\) is treated as a fixed spacetime smearing function, not
as a phase-space coordinate. Therefore
\begin{equation}
\mathcal L_{\delta_\chi}\Omega=0,
\tag{6.5}\label{eq:6.5}
\end{equation}
so \(\delta_\chi\) is a locally Hamiltonian vector field on the
unconstrained Carroll phase space, with Hamilton function \(Q_\chi\).
Two structural remarks are in order. First, the DBI-type action \eqref{eq:2.1} is of Nambu--Goto type. Its worldvolume geometry is induced, and after the Hamiltonian reduction it enters only through the multipliers \(N\), \(N^i\) and \(A_0\), which are not phase-space variables. Unlike the ILST null string discussed in Section~\ref{sec:null-string-comparison}, whose auxiliary field \(V^a\) carries the compensating transformation \(\delta_\chi V^a=-\chi V^a\), the canonical Carroll D\(_p\)-brane therefore possesses no independent worldvolume Carroll geometry on which a Weyl-type scaling could act. The transformation \eqref{eq:6.1}--\eqref{eq:6.3} is accordingly the matter-sector analogue of the null-string Carroll--Weyl symmetry. It extends the generator \(C_3=P\!\cdot\!X\) of the gauged null string \cite{Sheikh-Jabbari:2026tpf,Duary:2026lmk} to the transverse scalars, the tachyon and the worldvolume \(U(1)\) sector.
On the phase-space variables the two independent Carroll--Weyl scalings of the worldvolume Carroll structure \cite{Sheikh-Jabbari:2026vqh,Sheikh-Jabbari:2026tpf} induce the same transformation. They differ by the volume-modulating combination, which in this formulation leaves the matter variables inert and acts on the worldvolume frame alone, that is, on the multipliers \(N\), \(N^i\) and \(A_0\), while the criterion adopted here is formulated on the constraint algebra and is independent of the multipliers (see the footnote below). The transformation \eqref{eq:6.1}--\eqref{eq:6.3} is this common image, the matter-sector image of the volume-preserving combination, and it is the only transformation of this type available here. Second, the exclusion of the longitudinal pair \((X^\mu,P_\mu)\) from \eqref{eq:6.3} is not an arbitrary restriction. It is the most permissive choice, and it renders the kinetic terms \(P_\mu\partial_\tau X^\mu\), \(\pi^i\partial_\tau A_i\) and the term \(P_\mu\eta^{\mu\nu}P_\nu\) in the leading Hamiltonian constraints neutral under \(\chi\). Were \(X^\mu\) scaled as well, \(\delta_\chi(P_\mu\eta^{\mu\nu}P_\nu)=-2\chi P_\mu\eta^{\mu\nu}P_\nu\) would add a further obstruction with no compensating field available. The no-go result below is therefore proved for the variant of \(\chi\) most favourable to its survival.\footnote{For a global parameter the Gauss term transforms as \(\delta_\chi(A_0\mathcal G)=-\chi A_0\mathcal G\approx0\), so no transformation law for the multiplier \(A_0\) needs to be specified. The criterion is at the level of the constraint algebra and does not depend on the multipliers.}%

\subsection{Gauss obstruction}
\label{sec:gauss-obstruction}

Since
\(
\{\pi^i,Q_\chi\}=-\chi\pi^i,
\)
we find
\begin{equation}
\{G[\alpha],Q_\chi\}
=
-\int_\Sigma
\alpha
\Bigl[
(\partial_i\chi)\pi^i
+
\chi\,\partial_i\pi^i
\Bigr].
\tag{6.6}\label{eq:6.6}
\end{equation}
On the Gauss surface the second term vanishes,
\begin{equation}
\{G[\alpha],Q_\chi\}
\approx
-\int_\Sigma
\alpha\,(\partial_i\chi)\pi^i.
\tag{6.7}\label{eq:6.7}
\end{equation}
Local \(\chi\) is therefore obstructed by the Gauss constraint, whereas
global \(\chi\) preserves it.

\subsection{Diffeomorphism obstruction}
\label{sec:diffeo-obstruction}

For global \(\chi\),
\begin{equation}
\{H_i[\xi],Q_\chi\}=0,
\tag{6.8}\label{eq:6.8}
\end{equation}
because
\[
\delta_\chi(P_I\partial_i X^I)=0,
\quad
\delta_\chi(p_T\partial_i T)=0,
\quad
\delta_\chi(F_{ij}\pi^j)=0.
\]
For local \(\chi\), a direct computation gives
\begin{equation}
\begin{aligned}
\{H_i[\xi],Q_\chi\}
=
\int_\Sigma
\xi^i
\Bigl[
&(\partial_i\chi)P_I X^I
+
(\partial_i\chi)p_T T
\\
&+
(\partial_i\chi)A_j\pi^j
-
(\partial_j\chi)A_i\pi^j
\Bigr].
\end{aligned}
\tag{6.9}\label{eq:6.9}
\end{equation}
Local \(\chi\) is thus obstructed by the diffeomorphism constraint as well.

\subsection{Electric Hamiltonian obstruction}
\label{sec:electric-obstruction}

For global \(\chi\),
\(
\delta_\chi a_{ij}=2\chi\,a_{ij},
\
\delta_\chi\pi^i=-\chi\,\pi^i.
\)
The flux term is therefore invariant,
\begin{equation}
\delta_\chi(\pi^i a_{ij}\pi^j)=0,
\tag{6.10}\label{eq:chi-flux-invariant}
\end{equation}
while the potential term transforms as
\begin{equation}
\begin{split}
&\delta_\chi
\Bigl[
\tau_p^2 V(T)^2\det a_{ij}
\Bigr]
\\&=
\tau_p^2\det a_{ij}
\Bigl[
2p\,\chi V(T)^2
+
2\chi\,T V(T)V'(T)
\Bigr].
\end{split}
\tag{6.11}\label{eq:chi-electric-potential}
\end{equation}
Hence
\begin{equation}
\{H_E[\lambda],Q_\chi\}
=
\int_\Sigma
\lambda\,\tau_p^2\det a_{ij}
\Bigl[
2p\,\chi V^2
+
2\chi\,T V V'
\Bigr].
\tag{6.12}\label{eq:6.12}
\end{equation}
For a generic tachyon potential this expression is not identically zero,
and in general it is not weakly zero. Thus global \(\chi\) is obstructed by
the electric Hamiltonian constraint.

At the tachyon vacuum,
\(
V(T_{\min})=0,
\
V'(T_{\min})=0,
\)
the obstruction from the potential term vanishes. The vacuum condition
itself, however, is not preserved when \(T_{\min}\neq0\), since
\(
\delta_\chi T=\chi T.
\)
Thus \(\chi\) moves the system away from the tachyon vacuum unless it is
restricted not to act on \(T\), or unless \(T_{\min}=0\).

\subsection{Magnetic Hamiltonian obstruction}
\label{sec:magnetic-obstruction}

Similarly, for global \(\chi\), one has
\begin{equation}
\delta_\chi b_{ij}
=
2\chi\,a_{ij}+\chi\,F_{ij}.
\tag{6.13}\label{eq:6.13}
\end{equation}
The determinant therefore varies as
\begin{equation}
\delta_\chi\det b_{ij}
=
\chi\det b_{ij}\,
b^{ji}(2a_{ij}+F_{ij}).
\tag{6.14}\label{eq:6.14}
\end{equation}
Including the tachyon potential,
\begin{equation}
\begin{aligned}
\{H_M[\lambda],Q_\chi\}
=
\int_\Sigma
\lambda\,\tau_p^2
\Bigl[
&2\chi\,T V V'\det b_{ij}
\\
&+
\chi V^2\det b_{ij}\,
b^{ji}(2a_{ij}+F_{ij})
\Bigr].
\end{aligned}
\tag{6.15}\label{eq:6.15}
\end{equation}
For generic \(V(T)\), this expression is not identically zero. In the next
subsection we give an explicit on-shell point satisfying
\(\mathcal G=0\), \(\mathcal H_i=0\), and \(h_M=0\), for which it is nonzero.
At the tachyon vacuum, \(V=V'=0\), the obstruction vanishes, subject to the
same caveat: \(\chi\) does not preserve \(T=T_{\min}\) unless
\(T_{\min}=0\) or \(\chi\) is restricted.

\subsection{Explicit on-shell obstruction for arbitrary \(p\)}
\label{sec:magnetic-counterexample}

We now exhibit explicit points of the constraint surface, in both the
magnetic-like and the electric-like sectors, at which global \(\chi\) is
obstructed. They are generic-\(V\) points, not tachyon-vacuum points: they
assume \(V(0)=v\neq0\). Their purpose is to show that \(\chi\) is not a
symmetry of the full constrained phase space of the non-BPS Carroll
D\(_p\)-brane. The special tachyon-vacuum exception is discussed in
Section~\ref{sec:no-go}.

\subsubsection{The case \(p\ge2\)}

Assume \(p\ge2\) and that there are at least \(p-1\) transverse scalars,
i.e.
\[
D-p-1\ge p-1
\quad\Longleftrightarrow\quad
D\ge2p.
\]
Choose \(p-1\) transverse scalars and denote them locally by
\[
X^a,\qquad a=1,\ldots,p-1,
\]
where \(X^a\) are shorthand for selected transverse coordinates
\(X^{I_a}\). Set their gradients to
\[
\partial_i X^a=\delta_i^a,
\qquad
i=1,\ldots,p-1,
\qquad
a=1,\ldots,p-1,
\]
and take all other transverse scalars to be constant. Then
\begin{equation}
a_{ij}
=
\partial_iX^a\partial_jX^a
=
\operatorname{diag}(\underbrace{1,\ldots,1}_{p-1},0).
\tag{6.16}\label{eq:counter-aij-general}
\end{equation}

Choose a constant magnetic flux in the last two spatial directions:
\begin{equation}
F_{p-1,p}=f,
\qquad
F_{p,p-1}=-f,
\qquad
f\neq0,
\tag{6.17}\label{eq:counter-F-general}
\end{equation}
with all other components of \(F_{ij}\) vanishing. Then
\(
b_{ij}=a_{ij}+F_{ij}
\)
has block form
\begin{equation}
b
=
\begin{pmatrix}
I_{p-2} & 0 & 0\\
0 & 1 & f\\
0 & -f & 0
\end{pmatrix},
\qquad
\det b=f^2.
\tag{6.18}\label{eq:counter-b-general}
\end{equation}
Its inverse is
\begin{equation}
b^{-1}
=
\begin{pmatrix}
I_{p-2} & 0 & 0\\
0 & 0 & -\dfrac{1}{f}\\
0 & \dfrac{1}{f} & \dfrac{1}{f^2}
\end{pmatrix}.
\tag{6.19}\label{eq:counter-binv-general}
\end{equation}

Put the tachyon at
\begin{equation}
T=0,
\qquad
p_T=0,
\qquad
\partial_iT=0,
\tag{6.20}\label{eq:counter-tachyon-general}
\end{equation}
and assume
\(
V(0)=v\neq0.
\)
This is not the tachyon vacuum, but it is a legitimate point of the generic
non-BPS Carroll phase space.

Now we set the gauge and matter momenta as follows:
\begin{equation}
\pi^i=0,
\qquad
P_I=0
\tag{6.21}\label{eq:counter-momenta-general}
\end{equation}
for all transverse scalars $I$.
For the longitudinal sector, we set the spatial momenta to zero,
\begin{equation}
P_i=0,
\qquad
i=1,\ldots,p,
\tag{6.22}\label{eq:counter-Pi-long-general}
\end{equation}
and choose
\begin{equation}
P_0=\tau_p v|f|
\tag{6.23}\label{eq:counter-P0-general}
\end{equation}
Finally we choose
\begin{equation}
\partial_iX^0=0,
\qquad
\partial_iX^j=\delta_i^j.
\tag{6.24}\label{eq:counter-X-general}
\end{equation}

We verify that this point lies on the full constraint surface. The Gauss
constraint is satisfied because
\(
\mathcal G=\partial_i\pi^i=0.
\)
Each term of diffeomorphism density $H_i$ vanishes at the chosen point:
\[
P_\mu\partial_iX^\mu=0,
\quad
P_I\partial_iX^I=0,
\quad
p_T\partial_iT=0,
\quad
F_{ij}\pi^j=0,
\]
so \(\mathcal H_i=0.\)
The magnetic Hamiltonian constraint is
\[
h_M
=
-P_0^2+\tau_p^2v^2\det b.
\]
Using
\(
P_0=\tau_p v|f|,
\
\det b=f^2,
\)
we obtain
\(
h_M=0.
\)
Thus the point is on the full constraint surface.

We now evaluate the \(\chi\) obstruction. Since \(T=0\),
\[
\delta_\chi V(T)=\chi\,T V'(T)=0,
\]
so only the determinant variation contributes:
\begin{equation}
\{H_M[\lambda],Q_\chi\}
=
\int_\Sigma
\lambda\,\tau_p^2v^2
\chi\det b\,
b^{ji}(2a_{ij}+F_{ij}).
\tag{6.25}\label{eq:counter-obstruction-general}
\end{equation}
From the block form of \(b^{-1}\), one finds
\(
b^{ji}a_{ij}
=
p-2,
\)
and
\(
b^{ji}F_{ij}
=
2.
\)
Therefore
\[
b^{ji}(2a_{ij}+F_{ij})
=
2\left(p-2\right)
+2
=
2(p-1).
\]
Multiplying by \(\det b=f^2\), we get
\begin{equation}
\det b\,
b^{ji}(2a_{ij}+F_{ij})
=
2(p-1)f^2.
\tag{6.26}\label{eq:counter-product-general}
\end{equation}
Hence
\begin{equation}
\{H_M[\lambda],Q_\chi\}
=
\int_\Sigma
\lambda\,
2\chi\,\tau_p^2v^2
(p-1)f^2
\neq0,
\quad
p\ge2.
\tag{6.27}\label{eq:magnetic-counterexample-general}
\end{equation}

\subsubsection{The case \(p=1\)}

For \(p=1\) there is no nonzero spatial two-form \(F_{ij}\). Assume instead
that there is at least one transverse scalar, i.e. \(D\ge3\). We can choose
\begin{equation}
a_{11}=1,
\qquad
F_{11}=0,
\qquad
b_{11}=1.
\tag{6.28}\end{equation}
We can set
\begin{equation}
\begin{aligned}
  \pi^1 &= 0, & \quad P_I &= 0, & \quad P_0 &= \tau_p v, & \quad T &= 0, \\[6pt]
  p_T &= 0, & \quad \partial_1 T &= 0, & \quad V(0) &= v \neq 0.
\end{aligned}
\tag{6.29}
\end{equation}
We also take \(\partial_1X^0=0\) and \(P_1=0\).
Then
\(
\mathcal G=0,
\
\mathcal H_1=0,
\
h_M=-P_0^2+\tau_p^2v^2=0.
\)
But
\(
b^{ji}(2a_{ij}+F_{ij})=2,
\
\det b=1,
\)
so
\begin{equation}
\{H_M[\lambda],Q_\chi\}
=
\int_\Sigma
\lambda\,
2\chi\,\tau_p^2v^2
\neq0.
\tag{6.30}\label{eq:magnetic-counterexample-p1}
\end{equation}

\subsubsection{The electric-like sector}

The electric sector admits an equally explicit on-shell obstruction. Assume
\(D\ge 2p+1\), so that \(a_{ij}\) can be built from \(p\) transverse scalars
alone. Choose \(\partial_iX^a=\delta_i^a\) for \(a=1,\dots,p\), with all other
transverse scalars constant, so that \(a_{ij}=\delta_{ij}\) and \(\det a=1\).
Take the tachyon at the constant value \(T=0\) with \(p_T=0\),
\(\partial_iT=0\) and \(V(0)=v\neq0\), and set
\[
\pi^i=0,\qquad P_I=0,\qquad P_i=0,\qquad \partial_iX^0=0,\qquad
P_0=\tau_p v .
\]
Then \(\mathcal G=0\), \(\mathcal H_i=0\), and
\(h_E=-P_0^2+\tau_p^2v^2=0\), so the point lies on the full constraint
surface. As in the magnetic example, this is not the tachyon vacuum but a
legitimate point of the generic non-BPS Carroll phase space, since
\(V(0)=v\neq0\). Since \(T=0\), Eq.~\eqref{eq:6.12} gives
\[
\{H_E[\lambda],Q_\chi\}
=\int_\Sigma \lambda\;2p\,\chi\,\tau_p^2 v^2
\;\neq\;0 ,
\]
for any global (constant) \(\chi\neq0\) and \(\lambda\ge0\) not identically zero. If instead one
allows a linear tachyon profile, \(\partial_pT=1\) with the
\(\sigma\)-dependent longitudinal momentum \(P_0(\sigma)=\tau_p V(T(\sigma))\),
the same conclusion holds already for \(D\ge2p\).

\subsubsection{Conclusion of the counterexample}

Thus, for any \(p\ge2\) with \(D\ge2p\), and for \(p=1\) with \(D\ge3\), there
exists a point on the magnetic constraint surface at which global \(\chi\)
fails to preserve the Hamiltonian constraint. For \(p\ge2\) the obstruction is
proportional to
\(
(p-1)f^2,
\)
while for \(p=1\) it is proportional to \(1\). In both cases it is nonzero for \(V(0)=v\neq0\). This proves that global \(\chi\) is not a weak symmetry of the full magnetic constrained phase space.
The electric-like example establishes the same conclusion for the electric constrained phase space, for \(D\ge2p+1\).

\subsubsection{Robustness of the obstruction}
\label{sec:robust}

Three features make the obstruction structurally stable.
\emph{(i) Algebraic versus derivative obstructions.} In the ILST null string
the variation of the action under \eqref{eq:null-chi} reduces to
\(\kappa\int V^aV^b\,X\!\cdot\!\partial_aX\,\partial_b\chi\), which is
proportional to \(\partial\chi\) and is therefore removed by the profile
restriction \(V^a\partial_a\chi=0\). The leading brane obstructions
\eqref{eq:6.12} and \eqref{eq:6.15}, by contrast, are proportional to \(\chi\)
itself. No restriction on the spacetime profile of \(\chi\) can remove them at
nonzero \(\chi\). They vanish identically only in the singular cases identified in item (iii) below, and otherwise only pointwise, for instance at a tachyon vacuum with \(V=V'=0\).
\emph{(ii) Invariance under improvement.} If \(\varphi_a\) denote first-class
combinations of \(H_{E/M}\), \(H_i\) and \(\mathcal G\), then
\(\{H_E[\lambda],Q_\chi+\int c^a\varphi_a\}\approx\{H_E[\lambda],Q_\chi\}\)
for arbitrary coefficients \(c^a\). The obstruction class is therefore
invariant under any redefinition of \(Q_\chi\) by constraint terms, and no
such improvement makes \(Q_\chi\) first class.
\emph{(iii) Independence of the potential profile.} For a power-law potential
\(V\propto T^n\) one has \(pV^2+TVV'=(p+n)V^2\), so that \eqref{eq:6.12}
becomes \(2\chi(p+n)\tau_p^2V^2\det a=2\chi(p+n)\bigl(h_E-P_\mu\eta^{\mu\nu}P_\nu
-\pi^ia_{ij}\pi^j\bigr)\). Evaluated at the electric-like on-shell point exhibited above, which assumes \(V(0)=v\neq0\), this equals \(2\chi(p+n)P_0^2\), nonzero for \(n\neq-p\). For potentials that are not finite and nonzero at \(T=0\) the same evaluation is made at a constant-\(T\) on-shell point with \(T\neq0\), where \(P_0=\tau_pV(T)\sqrt{\det a}\). At \(n=-p\) the coefficient vanishes and \eqref{eq:6.12} is identically zero. The potential \(V\propto T^{-p}\) is singular at the value \(T=0\) at which the counterexamples sit, and for that family the electric-like sector does not obstruct global \(\chi\). The class of potentials for which the obstruction is established here is therefore the class that is finite and nonzero at \(T=0\), where the bracket multiplying \(\lambda\tau_p^{2}\det a\) in \eqref{eq:6.12} reduces to \(2p\chi V(0)^2\neq0\). Outside it, for \(V\propto T^{-p}\) the magnetic integrand of \eqref{eq:6.15} reduces to \(-\chi V^{2}\det b\,\mathrm{tr}(b^{-1}F)\), which is nonzero for \(p\ge2\) and generic \(F_{ij}\neq0\). At \(p=1\) one has \(F_{ij}\equiv0\), so \(b_{ij}=a_{ij}\) and \eqref{eq:6.15} vanishes identically for \(V\propto T^{-1}\), in parallel with \eqref{eq:6.12}.

\subsection{No-go statement for \(\chi\) as a first-class gauge generator}
\label{sec:no-go}

The results of the previous subsections can be summarized as follows.

\begin{quote}
\noindent\textbf{No-go statement.}\quad
Within the canonical Carroll symplectic structure and the first-class
constraint system
\[
H_{E/M}[\lambda],
\qquad
H_i[\xi],
\qquad
G[\alpha],
\]
the Carroll--Weyl transformation \(\chi\) cannot be promoted to a
first-class gauge generator of the constrained phase space, except in
trivial or highly restricted cases, the \(p=1\) member of the singular family \(V\propto T^{-p}\) of Section~\ref{sec:robust} being of the latter type.

Local \(\chi\) is obstructed by the Gauss constraint
\eqref{eq:6.7} and by the diffeomorphism constraint
\eqref{eq:6.9}. For a generic tachyon potential \(V(T)\),
global \(\chi\) is obstructed by the Hamiltonian constraint in both the
electric-like and magnetic-like sectors, as shown in
\eqref{eq:6.12} and \eqref{eq:6.15}. The obstruction is nonzero on the constraint surface in both sectors, as demonstrated by the explicit on-shell examples of Section~\ref{sec:magnetic-counterexample}.

At a generic tachyon vacuum with \(T_{\min}\neq0\), even if
\(V(T_{\min})=V'(T_{\min})=0\), the transformation
\(\delta_\chi T=\chi T\) does not preserve the vacuum condition
\(T=T_{\min}\). The only nontrivial exception relevant to the vacuum sector
is the special case
\begin{equation*}
\begin{split}
T_{\min}&=0,
\quad
V(0)=V'(0)=0,\\
T&=p_T=0,
\quad
\partial_i\chi=0.
\end{split}
\end{equation*}
In this case \(\chi\) preserves the vacuum conditions and the leading
Hamiltonian constraints, but it is at most a global symmetry of the vacuum
sector. It is not a local gauge redundancy, and \(Q_\chi\) is not itself a
first-class constraint.

Therefore, except for this global \(T_{\min}=0\) vacuum-sector exception, the singular potential \(V\propto T^{-1}\) at \(p=1\) identified in item (iii) above, and other trivial restrictions, \(\chi\) does not reduce the physical phase space. It is a formal grading of the unconstrained Carroll matter sector
rather than a gauge symmetry of the constrained system.
\end{quote}

\subsection{Comparison with null strings}
\label{sec:null-string-comparison}

The status of \(\chi\) in the Carroll D$_p$-brane system should be contrasted
with its role in null string theory. In the ILST formulation
\cite{Isberg:1993av}, the null string action is
\begin{equation}
S_{\mathrm{null}}
=
\frac{\kappa}{2}
\int d^2\sigma\,
V^aV^b\,\partial_a X^\mu\,\partial_b X_\mu,
\tag{6.31}\label{eq:null-action}
\end{equation}
and it possesses a restricted local \(\chi\)-symmetry
\begin{equation}
\delta_\chi X^\mu=\chi X^\mu,
\qquad
\delta_\chi V^a=-\chi V^a,
\qquad
V^a\partial_a\chi=0.
\tag{6.32}\label{eq:null-chi}
\end{equation}
This restricted symmetry has been argued to require an additional
constraint and to change the local degree-of-freedom count from the
commonly quoted \(D-2\) to \(D-3\) \cite{Sheikh-Jabbari:2026cnj}. In the
Carroll--Weyl gauged formulation, the symmetry is completed by a third
first-class constraint
\begin{equation}
C_3=P\cdot X,
\tag{6.33}\label{eq:C3}
\end{equation}
and the BRST analysis treats \(\chi\) as an additional first-class gauge
generator
\cite{Sheikh-Jabbari:2026tpf,Duary:2026lmk}. Related path-integral and quantum analyses appear
in \cite{Duary:2026rlo,Lindstrom:2026zno,Figueroa-OFarrill:2025njv,Chen:2026cau}.

The D$_p$-brane case differs for three structural reasons. First, and most fundamentally, the worldvolume geometry of the DBI action is induced rather than auxiliary. In the canonical formulation the lapse, the shift and \(A_0\) enter only as multipliers, so there is no worldvolume Carroll geometry carrying an independent Weyl-type scaling. This is in contrast to the null string, where the auxiliary field \(V^a\) transforms under \(\chi\), and where the gauged formulation promotes the two Carroll--Weyl scalings of the worldsheet Carroll geometry to local symmetries \cite{Sheikh-Jabbari:2026vqh,Sheikh-Jabbari:2026tpf}.
On the brane neither of the two survives as an independent geometric scaling, since they differ only through the worldvolume frame. The matter phase space therefore sees a single transformation, the one analysed in Section~\ref{sec:chi-definition}. The compensating transformation \(\delta_\chi V^a=-\chi V^a\) is what allows the restricted \(\chi\)-symmetry to survive there. On the brane, no multiplier transformation can compensate the variation of the DBI potential, and in any case the canonical criterion is the first-class property of \(Q_\chi\), which is independent of the multipliers. Second, the phase
space contains a worldvolume \(U(1)\) sector with the Gauss constraint
\(
\mathcal G=\partial_i\pi^i.
\)
Since \(\pi^i\) transforms under diffeomorphisms only modulo
\(\mathcal G\), and since local \(\chi\) changes \(\pi^i\), the Gauss
constraint obstructs local \(\chi\) unless \(\chi\) is constant or the
electric flux is restricted. Third, the non-BPS D$_p$-brane contains the
tachyon potential \(V(T)\). The transformation \(\delta_\chi T=\chi T\) is
generally incompatible with a nonzero vacuum value \(T_{\min}\), and the
Hamiltonian obstruction contains terms proportional to
\(
V(T)^2,\
T V(T)V'(T).
\)
Only for \(T_{\min}=0\), \(V(0)=V'(0)=0\), with the vacuum-sector conditions
\(T=p_T=0\) imposed and \(\chi\) restricted to be global, does the vacuum
sector avoid the Hamiltonian obstruction.

\emph{Why a connection cannot supply the missing compensation.} A Carroll--Weyl connection enters the action through covariant derivatives, so its contribution to the variation is proportional to \(\partial\chi\), and it reduces to a total divergence upon integration by parts. The obstructions that survive for global \(\chi\), \(\Theta_{\mathrm{alg}}=\tau_p^2\det a\,[2pV^2+2TVV']\) in \eqref{eq:6.12} and its magnetic analogue in \eqref{eq:6.15}, are not of this type. For generic \(V\) the first of them is a nonvanishing function of \(T\) times \(\det a\), so it cannot be written as the divergence of a local current built from the fields, and no local connection term cancels it. The null string is different in precisely this respect, its algebraic part vanishing and its obstruction being proportional to \(\partial\chi\), which is what allows the compensating transformation \(\delta_\chi V^a=-\chi V^a\) of a field of weight \(-1\) to absorb it. The missing ingredient here is therefore not a connection but a compensator of that weight, and introducing one dynamically yields a Weyl-invariant completion in which the present theory is a gauge slice, that is, a different theory. The tachyon itself supplies that weight only for \(V\propto T^{-p}\), the singular family of item (iii) above, whose potential term is weight-neutral on its own. Within the present field content no connection promotes \(\chi\) to a first-class generator, and whether such a completion exists and is anomaly free is left open.

The original construction in \cite{Kluson:2017fam} provides the canonical Carroll non-BPS D$_p$-brane actions and their tachyon-vacuum solutions, including the interpretation of the vacuum as a gas of Carroll strings under suitable restrictions \cite{Kluson:2017fam}. Related work constructs D-branes in Carrollian backgrounds and studies tachyon kinks and vacuum solutions in generalized Carroll space-times \cite{Kluson:2022jxh}. These works establish the dynamical and solution-space context of Carroll non-BPS D$_p$-branes, but they do not formulate the Carroll--Weyl transformation as a first-class
gauge generator.

Thus the present result clarifies the status of the Carroll--Weyl
transformation in Carroll D-brane systems. Unlike the null string, where \(\chi\) can be completed to a genuine first-class gauge generator, in the canonical non-BPS
Carroll D$_p$-brane \(\chi\) is at most a formal grading or a restricted
global symmetry, and it does not reduce the physical phase space. The
mechanism by which a restricted null-string \(\chi\)-symmetry is completed
to a third first-class constraint does not automatically extend to the
worldvolume \(U(1)\) and tachyon sectors of the D$_p$-brane.

\section{Conclusion}
\label{sec:conclusion}

In this paper we have analyzed the covariant phase space of Kluso\v{n}'s
canonical Carroll non-BPS D\(_p\)-brane actions. Section~\ref{sec:actions}
identified the parent model and distinguished the two leading sectors of
Kluso\v{n}'s construction: the electric-like sector, in which the gauge field
is scaled together with the longitudinal coordinates and the leading
Hamiltonian density contains the flux term \(\pi^i a_{ij}\pi^j\), and the
magnetic-like sector, in which the gauge field is not scaled and the leading
Hamiltonian density contains \(\det(a_{ij}+F_{ij})\).
Sections~\ref{sec:actions} and
\ref{sec:phase-space} showed that the Carroll limit is well defined at
the symplectic level, because the canonical one-form is invariant under the
scaling of conjugate pairs. The difference between the two sectors lies in the
leading Hamiltonian constraint. In particular, \(\pi^i\) transforms as a
vector density only modulo the Gauss constraint, which makes the electric-like
constraint algebra weakly closed, whereas the magnetic-like Hamiltonian
brackets close strongly except for the diffeomorphism--diffeomorphism bracket.
Section~\ref{sec:degrees} separated the generic non-BPS sector from the
tachyon-vacuum sector: the generic sector carries \(D-1\) local phase-space
degrees of freedom, while the vacuum sector carries \(D-2\) only after
imposing second-class background conditions. We also clarified the electric
rank condition, which requires \(D\ge2p\) for a dynamical tachyon and
\(D\ge2p+1\) in the tachyon-vacuum effective sector.
Section~\ref{sec:chi} analyzed the Carroll--Weyl transformation \(\chi\).
Although \(\chi\) preserves the Carroll symplectic form and admits a charge,
it is obstructed by the Gauss, diffeomorphism, and Hamiltonian constraints.
The explicit on-shell counterexamples in the magnetic-like and electric-like
sectors show that global \(\chi\) is not a weak symmetry of the full
constrained phase space. The obstruction is algebraic in \(\chi\) and invariant
under improvement of \(Q_\chi\) by first-class terms, so it cannot be removed by
restricting the profile of \(\chi\) or by redefining the generator. The nontrivial exceptions are the highly restricted vacuum case \(T_{\min}=0\), \(V(0)=V'(0)=0\), with global \(\chi\) and the vacuum conditions \(T=p_T=0\) imposed, and the singular potential \(V\propto T^{-1}\) at \(p=1\) (Section~\ref{sec:robust}).

In summary, the main result is that the Carroll limit of Kluso\v{n}'s
non-BPS D\(_p\)-brane preserves the canonical symplectic structure but changes
the constraint surface. The electric-like and magnetic-like sectors differ by
the leading terms retained in the Hamiltonian constraint, and this difference
controls the weak or strong closure of the constraint algebra. The
Carroll--Weyl transformation \(\chi\) should therefore be interpreted as a
formal grading of the unconstrained matter sector, not as a first-class gauge
generator of the constrained system. This contrasts with null string theory,
where a restricted \(\chi\)-symmetry can be completed to an additional
first-class constraint and modify the physical degree-of-freedom count.

It would be interesting to extend this analysis in several directions.
First, one may study Carroll limits of D-branes with non-Abelian worldvolume
gauge fields \cite{Polchinski:1995mt,Myers:2000in,Seiberg:1999vs}, where the Gauss constraint is replaced by a non-Abelian one and
the obstruction structure of \(\chi\) may become richer. Second, it would be
useful to compare the present phase-space contraction with other Carroll
contraction schemes, including deformed light-cone and Kaluza--Klein-like null
reductions~\cite{Duval:2014uoa,Zeng:2026qtt}. Third, boundary phase spaces and Carrollian conformal symmetries
provide a natural setting for studying charges, edge modes, and possible
central extensions~\cite{Bagchi:2024qsb,Adami:2020ugu,Adami:2021nnf}. Finally, quantization of the constrained Carroll D\(_p\)-brane phase space, including BRST or Dirac-bracket quantization of the tachyon-vacuum sector, is an interesting open problem. It would also be interesting to ask whether a genuine geometric Carroll--Weyl symmetry can be realized on a Polyakov-type Carroll D-brane, in which an independent worldvolume Carroll geometry is introduced before the limit, along the lines of the gauged null-string construction \cite{Sheikh-Jabbari:2026vqh,Sheikh-Jabbari:2026tpf}. The present no-go result applies to the Nambu--Goto-type canonical theory, where the induced worldvolume geometry leaves only the matter-sector image of the volume-preserving Carroll--Weyl scaling available (Section~\ref{sec:chi-definition}). The question of a geometric realisation is not settled here. A constructive programme follows from this analysis. A \(\chi\)-symmetric D-brane would have to supply a field carrying the compensating weight and an anomaly analysis, and the present no-go identifies the obstructions that such a construction must evade.

\begin{acknowledgements}

Limin Zeng would like to thank only Yue Liao.

\end{acknowledgements}

\appendix
\section{Derivation of \(\{\pi^i,H_i[\xi]\}\)}
\label{app:A}

The only term in \(H_i[\xi]\) that depends on \(A_i\) is
\begin{equation}
\int_\Sigma
\xi^aF_{ab}\pi^b.
\label{eq:app-gauge-term}
\end{equation}
Its variation is
\begin{equation}
\delta\int_\Sigma \xi^aF_{ab}\pi^b
=
\int_\Sigma
\xi^a\pi^b
\Bigl(
\partial_a\delta A_b-\partial_b\delta A_a
\Bigr).
\label{eq:app-variation}
\end{equation}
Integrating by parts,
\begin{equation}
\frac{\delta H_i[\xi]}{\delta A_i}
=
-\partial_a(\xi^a\pi^i)
+
\partial_b(\xi^i\pi^b).
\label{eq:app-functional-derivative}
\end{equation}
Since
\begin{equation}
\{A_i(x),\pi^j(y)\}=\delta_i^j\,\delta(x-y),
\label{eq:app-Poisson}
\end{equation}
we have
\begin{equation}
\{\pi^i,H_i[\xi]\}
=
-\frac{\delta H_i[\xi]}{\delta A_i}
=
\partial_a(\xi^a\pi^i)
-
\partial_b(\xi^i\pi^b).
\label{eq:app-pi-H}
\end{equation}
Therefore
\begin{equation}
\{\pi^i,H_i[\xi]\}
=
\xi^a\partial_a\pi^i
-
\pi^b\partial_b\xi^i
+
\pi^i\partial_a\xi^a
-
\xi^i\partial_b\pi^b.
\label{eq:app-expanded}
\end{equation}
The first three terms combine into \(\mathcal L_\xi\pi^i\), and the last
term is \(-\xi^i\mathcal G\). Hence
\begin{equation}
\{\pi^i,H_i[\xi]\}
=
\mathcal L_\xi\pi^i-\xi^i\mathcal G.
\label{eq:app-pi-diffeo}
\end{equation}

\bibliographystyle{spphys}       
\bibliography{ex}   

\end{document}